\documentclass[12pt]{article}

\usepackage{cite,graphicx,amsmath,amssymb,bbm}
\usepackage{epsf,pifont,amsfonts,mathrsfs,floatflt}

\DeclareMathOperator{\tr}{tr} 
\newcommand{\be}{\begin{equation}}
\newcommand{\ee}{\end{equation}}
\newcommand{\bea}{\begin{eqnarray}}
\newcommand{\eea}{\end{eqnarray}}

\newcommand{\IR}{{\rm \text{\tiny IR}}}
\renewcommand{\Re}{\operatorname{Re}}
\renewcommand{\Im}{\operatorname{Im}}

\begin{document}

\thispagestyle{empty}
\vspace*{.2cm}
\noindent
HD-THEP-07-12 \hfill 24 May 2007

\vspace*{1.5cm}

\begin{center}
{\Large\bf Energy Transfer between Throats \\[.4cm] from a 10d
Perspective}\\[2.5cm]

{\large B.~v.~Harling, A.~Hebecker, and T.~Noguchi}\\[.5cm] {\it
Institut f\"ur Theoretische Physik, Universit\"at Heidelberg,
Philosophenweg 16 und 19, D-69120 Heidelberg, Germany} \\[.5cm]
{\small\tt (\,harling, a.hebecker,} {\small and} {\small\tt
t.noguchi@thphys.uni-heidelberg.de)} \\[2.0cm]

{\bf Abstract}
\end{center} 
\noindent 
Strongly warped regions, also known as throats, are a common feature
of the type IIB string theory landscape. If one of the throats is
heated during cosmological evolution, the energy is subsequently
transferred to other throats or to massless fields in the unwarped
bulk of the Calabi-Yau orientifold. This energy transfer proceeds
either by Hawking radiation from the black hole horizon in the heated
throat or, at later times, by the decay of throat-localized
Kaluza-Klein states. In both cases, we calculate in a 10d setup the
energy transfer rate (respectively decay rate) as a function of the
AdS scales of the throats and of their relative distance.  Compared to
existing results based on 5d models, we find a significant suppression
of the energy transfer rates if the size of the embedding Calabi-Yau
orientifold is much larger than the AdS radii of the throats.  This
effect can be partially compensated by a small distance between the
throats. These results are relevant, e.g., for the analysis of
reheating after brane inflation. Our calculation employs the dual
gauge theory picture in which each throat is described by a strongly
coupled 4d gauge theory, the degrees of freedom of which are localized
at a certain position in the compact space.

\newpage
\section{Introduction}
Strongly warped regions or throats are a common feature of the
landscape of type IIB string theory. More specifically, local
geometries which are similar to the Klebanov-Strassler
throat~\cite{KS} arise naturally in flux
compactifications~\cite{Giddings:2001yu} (see also~\cite{Verlinde})
and the distribution of vacua favours geometries with dynamically
generated large hierarchies~\cite{Denef:2004ze}. Under certain
assumptions, this can even be turned into a prediction for the
statistical distribution of multi-throat
configurations~\cite{Hebecker:2006bn}.

Multi-throat compactifications have been considered earlier~\cite{
smallnumbers,Cacciapaglia:2006tg} on the basis of the simpler
Randall-Sundrum model~\cite{RS}, which realizes the essential features
of the Klebanov-Strassler throat in a 5d geometry. Cosmological
implications of the energy transfer between throats have been studied
by a number of authors~\cite{Barnaby:2004gg,Kofman:2005yz,Frey:2005jk,
Chialva:2005zy,CT,Firouzjahi:2005qs,Langfelder}. An important
motivation for the analysis of cosmologies with heated throats comes
from the possibility of realizing brane inflation in the strongly
warped region of the compact manifold~\cite{Kachru:2003sx}.

In the present paper, we focus on the energy transfer between
different throats in a given type IIB compactification. If one of the
throats is heated during cosmological evolution, the energy is
subsequently transferred to other throats or to massless fields in the
unwarped bulk of the Calabi-Yau orientifold. This energy transfer
proceeds in two ways. If the temperature in a given throat is high
enough, it develops a black hole horizon~\cite{WGKP,GHKT} and energy
is lost by Hawking radiation. When the temperature drops below a
critical temperature $T_c$, a finite throat undergoes a phase
transition during which the black hole horizon is replaced by the
infrared cutoff region of the throat~\cite{Arkani-Hamed:2000ds,GHKT,
Creminelli:2001th}. Subsequently, the throat sector contains a
non-relativistic gas of Kaluza-Klein (KK) modes which decay to other
throats in the Calabi-Yau orientifold.

In both cases, we calculate the energy transfer rate (respectively
decay rate) as a function of the AdS scales of the throats and of
their relative distance. For the decay rate, we also demonstrate how
to determine its dependence on angular quantum numbers of the decaying
KK modes.  Moreover, we extend the analysis of~\cite{smallnumbers} to
a genuine 10d setup (for earlier related work
see~\cite{CT,Firouzjahi:2005qs}). To this end, we consider two
AdS$_5\times$S$^5$ throats embedded in a 6-dimensional torus. This is
a simplified model, but we argue that our results remain
parametrically correct also for more general geometries. As compared
to~\cite{smallnumbers}, we find a significant suppression of the
energy transfer rates if the size of the embedding Calabi-Yau
orientifold is much larger than the AdS radii of the throats. This
effect can be partially compensated by a small distance between the
throats. These results are relevant, e.g., for the analysis of
reheating after brane inflation.

It has been shown in~\cite{K,GKT} that the absorption cross sections
for scalars and transversely polarized gravitons by an
AdS$_5\times$S$^5$ throat agree with those by a stack of $N$ D3-branes
(for appropriate $N$). The fact that this agreement is exact in spite
of the use of leading-order perturbation theory in the strongly
coupled regime on the gauge theory side is explained by a
non-renormalization theorem~\cite{GK1}. Motivated by these results, we
employ the dual gauge theory picture in which each throat is described
by the world-volume gauge theory on the corresponding stack of
D3-branes. The world-volume theories on different D3-brane stacks are
coupled by the supergravity fields in the embedding manifold. The
decay and energy transfer rates then follow from the appropriate
quantum-field-theory tree-level diagrams. This calculation is
considerably simpler than the corresponding analysis in the gravity
picture, where one has to solve multi-dimensional tunneling problems.

We will use the above equivalence of the gravity and gauge theory
picture also for non-zero temperature, where the non-renormalization
theorem is violated. However, as we will show, this only leads to
${\cal O}(1)$ uncertainties. The same is true for the generalization
to the Klebanov-Strassler (approximate AdS$_5\times$T$^{1,1}$)
throat. Given that we are anyway ignorant about the detailed geometry
of the bulk space and of the specific throats which may appear in
realistic models, we can tolerate this uncertainty.

We emphasize that, although we refer to throats and the corresponding
large-$N$ D-brane stacks throughout the text, our results also apply
to stacks of fewer branes. This may be useful for the analysis of the
cosmology of a standard model which resides on D-branes in the
Calabi-Yau orientifold and heats up the surrounding throats.

\begin{figure}[t]\label{TwoThroats}
\begin{center}
\includegraphics[scale=0.5]{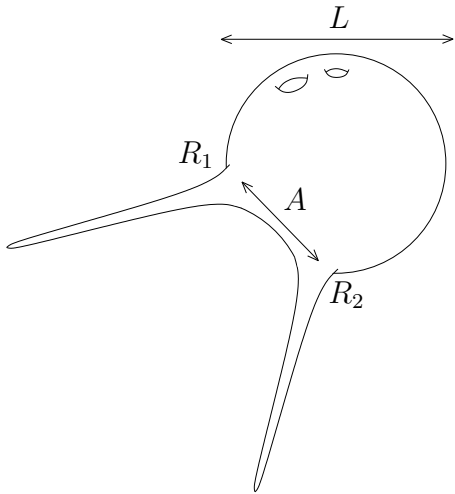} \put(-105,125){$R_1$}
\put(-48,72){$R_2$} \put(-65,107){$A$} \put(-48,176){$L$}
\caption{Two throats with radii $R_1$ and $R_2$ separated by a
distance $A$ inside a Calabi-Yau orientifold of total size $L$.}
\end{center}
\end{figure}

Our paper is organized as follows. In Sect.~\ref{energytransfer}, we
derive the energy loss of a heated throat to another throat which is
separated from the first one by a certain distance $A$
(cf. Fig.~\ref{TwoThroats}). This calculation is performed by
modelling both throats by stacks of D3-branes and replacing the
compact space by a torus. It is then straightforward to derive the
energy transfer rate by summing over the contributions of bulk KK
modes coupling to both throats. The resulting parametric behaviour
$\sim 1/A^8$ of the leading term remains valid for more general 6d
compact spaces and for more complicated throat geometries.

Sect.~\ref{KKdecay} describes an analogous calculation for the decay
rate of KK modes localized in one throat to fields in a distant
throat.  In the gauge theory picture, the decaying KK modes are
represented by glueballs.  Thus, we first derive the effective vertex
for the coupling of these glueballs to bulk fields. After that, the
calculation proceeds analogously to that in the previous
section. Finally, we compare certain limiting cases of our result with
calculations in the gravity picture and with formulae from the
literature.

Our conclusions are given in Sect.~\ref{conclusions}, where we also
outline possible applications of our results. A relevant integral is
evaluated in the Appendix.

\section{Energy transfer between two throats}\label{energytransfer}
\begin{figure}[ht]\label{scat}
\begin{center}
\includegraphics[scale=0.5]{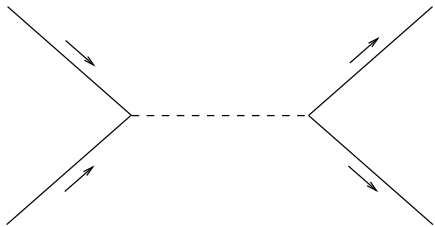}
\caption{Feynman diagram for the scattering of fields on one brane
stack into fields on another brane stack.}
\end{center}
\end{figure}

Let us consider a compactification manifold containing two throats,
one of which is heated to a certain temperature $T$. An interesting
quantity for cosmology is the rate of energy transfer to the other
throat. In the following, we will determine this rate using the
description of the throats in terms of D-brane stacks. In this
picture, a heated throat corresponds to a heated world-volume gauge
theory. The world-volume theories on the two brane stacks are coupled
by the supergravity fields in the embedding space.  Thus, energy
transfer between the two throats is, in this picture, due to processes
of the type shown in Fig.~\ref{scat}, where fields in the thermal
plasma on one brane stack scatter into fields on the other brane
stack.
 
We will perform the corresponding calculation for a simple example --
two semi-infinite AdS$_5 \times$S$^5$ throats embedded in a
6-dimensional torus of uniform size $L$. These throats are the
near-horizon geometries of black 3-branes, which in turn correspond to
stacks of D3-branes (see e.g.~\cite{Verlinde,Chan:2000ms}). For each
throat, the S$^5$ radius $R$ is related to the D-brane number $N$ by
\begin{equation}
\label{RN}
R^4 = \frac{\kappa_{10} N}{2 \pi^{5/2}}\,.
\end{equation}
As it stands, this is not a consistent compactification since
negative-charge objects are needed to absorb the flux of the
branes. However, in the course of our calculation we will argue that
including these and other objects (e.g.  further D-branes) as well as
using a different embedding manifold and a different throat geometry
only leads to $\mathcal{O}(1)$ corrections.

We will restrict our calculation to the mediation by the dilaton, the
Ramond-Ramond scalar and the graviton polarized parallel to the
branes.  In the gravity picture these three fields satisfy the same
wave equation~\cite{GKT}. Correspondingly, in the gauge theory picture
their effect in mediating energy transfer is parametrically the
same.\footnote{ This can also be inferred from the relevant part of
the DBI action, which couples them to the world-volume theories on the
D3-branes.  } Hence, we can further restrict our calculation to one of
the three fields, which we take to be the dilaton. In particular, we
will not consider the effect of fermions living in the embedding
manifold. In fact, in~\cite{ Hosomichi:1998ei} the absorption cross
section of dilatinos by 3-branes was calculated and found to agree
with the result for the dilaton. Therefore, we expect the fermions to
give parametrically the same contribution as the fields that we
consider.

The (low-energy) world-volume theory on $N$ parallel D3-branes is
$\smash{\mathcal{N}=4}$ U($N$) super Yang-Mills. Its field content is
given by the field strength $F_{\alpha \beta}$ in the adjoint
representation, six adjoint scalars $X^i$ corresponding to the
positions of the branes, and fermionic superpartners. The coupling
between the dilaton and the field strength follows from the standard
10d supergravity action with a stack of D3-branes (see
e.g.~\cite{Aharony:1999ti}) 
\be
\label{action}
S=\frac{1}{2\kappa_{10}^2}\int d^{10}x\sqrt{g}\left[{\cal
R}-\frac{1}{2} (\partial\phi)^2+\cdots\right]+\int
d^4x\left[-\frac{1}{4}e^{-\phi}\mbox{tr}
F_{\alpha\beta}^2+\cdots\right]\,, 
\ee 
where, here and below, we work
in the 10d Einstein frame. We ignore couplings to fermions, since they
are proportional to the fermionic equations of motion and thus give no
contributions to S-matrix elements~\cite{GKT}. Direct couplings
between the dilaton $\phi$ and the scalars $X^i$ are absent.
Canonically normalizing the dilaton kinetic term and allowing for
brane fluctuations, we get~\cite{K}
\begin{equation}
\label{coupling2}
S \supset \frac{\kappa_{10}}{2^{3/2}}\left[ \int\! d^{4}x \, \phi(x,
\langle \vec{X} \rangle) \, \tr F_{\alpha \beta}^2 + \sum_l \int\!
d^{4}x \frac{\kappa_{10}^{l/2} }{ l!\pi^{l/4} } \left(\partial_{i_1}
\cdots \partial_{i_l} \phi \right) \tr \left( X^{i_1} \cdots X^{i_l}
\, F_{\alpha \beta}^2 \right)\right]\!,
\end{equation}
where $\smash{\langle \vec{X} \rangle}$ is the position of the brane
stack.  The $X^i$ are also defined such that their kinetic terms are
canonically normalized. As can be seen from Eq.~\eqref{coupling2},
couplings involving the $X^i$ as well as $F_{\alpha \beta}$ are
suppressed by extra factors of $\smash{\kappa_{10}^{l/2}}$ and can
therefore be ignored.

\subsection{Energy loss rate to flat 10d space}

Before we proceed, we should check whether a calculation in terms of
weakly coupled gauge fields is a good approximation in the strongly
coupled regime of the gauge theory. At zero temperature, this is
adequate due to the non-renormalization theorem derived
in~\cite{GK1}. However, the gauge theory is at finite temperature,
which breaks supersymmetry. With supersymmetry being broken, the
non-renormalization theorem from~\cite{GK1} cannot be expected to hold
and it is not immediately clear why to trust our
calculation. Therefore, we analyse a simple example in both the gauge
theory and the gravity picture and compare the results. Namely, we
consider a heated stack of D3-branes in flat 10d space which is dual
to a non-extremal black 3-brane and calculate the energy loss rate in
both pictures.

We model the heated, strongly-coupled gauge theory on the D3-brane
stack by a thermal plasma of free fields. In principle, one would have
to use finite temperature field theory for the calculation of the
energy loss rate. However, as we are only interested in the correct
order of magnitude, we can perform a zero-temperature calculation
using a thermal particle distribution in the initial state. Following
from Eq.~\eqref{coupling2}, the cross section for scattering of two
gauge bosons into one dilaton is 
\be
\label{csb}
\sigma \sim \kappa_{10}^2 \, s^3 
\ee 
up to $\mathcal{O}(1)$
prefactors, where $\smash{\sqrt{s}}$ is the energy of the gauge bosons
in the center of mass frame. From Eq.~\eqref{csb}, we can calculate
the rate of energy loss per world-volume of the branes induced by this
scattering process. This is done by thermally averaging the product of
cross section and lost energy, in analogy to the standard calculations
of reaction rates in a hot plasma~\cite{ swo,HMR}:
\begin{equation}
\label{rate}
\dot{\rho} = \frac{1}{2} \int d^3k_1 \, d^3k_2 \; f(\omega_1) \,
f(\omega_2) \, \sigma v \, (\omega_1 + \omega_2)\,.
\end{equation}
Here
\begin{equation}
f(\omega) = \frac{1}{4 \pi^3(e^{\omega/T} -1)}
\end{equation}
is the distribution function for the gauge bosons, $v$ is the relative
velocity of the colliding particles, and $T$ is the temperature of the
heated gauge theory. Inserting Eq.~\eqref{csb} into Eq.~\eqref{rate},
we get the energy loss rate due to scattering of one gauge boson
species. To get the total energy loss rate, we have to sum over all
species and polarizations.  In a U($N$) gauge theory there are $N^2$
gauge bosons. Thus, there is an extra factor of $2 N^2$ coming from
the summation. Using Eq.~\eqref{RN} and neglecting prefactors of order
one coming from the integration in Eq.~\eqref{rate}, we get 
\be
\label{sbel2}
\dot{\rho} \sim R^8 \, T^{13}, 
\ee 
where $R$ is the AdS scale of the
corresponding black 3-brane.

Energy loss from the non-extremal black 3-brane is due to Hawking
radiation emitted by its black hole horizon. The corresponding rate
per brane world-volume $\dot{\rho}$ is given by a generalization of
the Hawking formula (see e.g.~\cite{Aharony:1999ti}). If we restrict
ourselves to the dilaton, we get 
\be 
\dot{\rho} = \int \frac{d^9k}{(2
\pi)^9} \, \frac{ v \, \omega \, \sigma_T(\omega)}{e^{\omega /T} -1},
\ee 
where $v$ is the velocity of the emitted particles and $T$ is the
Hawking temperature of the horizon. The absorption cross section
$\sigma_T(\omega)$ of a dilaton by a non-extremal black 3-brane was
calculated in~\cite{Satoh:1998ss}. The result is $\sigma_T(\omega)=
\sigma_0(\omega) \, f(\omega/T)$, where $\omega$ is the energy of the
incident dilaton, $f$ is some function of the dimensionless ratio
$\omega/T$, and $\sigma_0(\omega) \sim \omega^3 R^8$ is the absorption
cross section by an extremal black 3-brane with AdS scale $R$ which
was already determined in~\cite{K}.  Inserting $\sigma_T(\omega)$ and
performing the integral, we get 
\be
\label{sbel1}
\dot{\rho} \sim R^8 \, T^{13}.  
\ee 
Here we have neglected prefactors
of order one which come in particular from the integration over
$f(\omega/T)$.

Both results for the energy loss rate, Eqs.~\eqref{sbel2} and
\eqref{sbel1}, agree up to $\mathcal{O}(1)$ factors. Accordingly, a
weak-coupling calculation in the gauge theory picture gives the right
order of magnitude.  The crucial ingredient is the fact that the
absorption cross section $\sigma_T$ of a dilaton by a non-extremal
black 3-brane differs from the zero-temperature absorption cross
section $\sigma_0$ only by a function of $\lambda \equiv \omega/T$. By
gauge/gravity duality, this means that the gauge boson-dilaton vertex
is corrected by a function of $\lambda$ at non-zero
temperature.\footnote{ This is also the case if one takes
finite-temperature effects properly into account on the gauge theory
side.  }  Accordingly, the cross section for the process in
Fig.~\ref{scat} that we will calculate assuming weak coupling and zero
temperature has to be corrected by a function of $\lambda$. However,
inserting the corrected cross section into Eq.~\eqref{rate} and
performing the integral will just give a different ${\cal O}(1)$
prefactor, which we ignore anyway.

\subsection{Energy transfer rate to a different throat}
Let us now calculate the cross section for the process in
Fig.~\ref{scat}.  To this end, we need the KK expansion of the dilaton
in a 6d torus,
\begin{equation}
\label{expansion}
\phi(x,\langle \vec{X} \rangle) = \sum_{\vec{n} \in \mathbb{Z}^6}
\frac{1} {L^3} \, e^{2 \pi i \vec{n} \langle \vec{X} \rangle /L} \,
\Phi_{\vec{n}}(x),
\end{equation}
where $L$ is the size of the torus and the expression is already
evaluated at the position $\smash{\langle \vec{X} \rangle}$ of one
brane stack. The mass of the $\vec{n}$th KK mode is
$\smash{m_{\vec{n}}=2 \pi |\vec{n}|/L}$.  Inserting
Eq.~\eqref{expansion} into Eq.~\eqref{coupling2} and using
$\smash{\kappa_{10} = M_{10}^{-4}}$, one sees that the vertex for the
$\vec{n}$th KK mode in Fig.~\ref{scat} is
\begin{equation}
\label{vertex}
\sim\frac{s}{M_{10}^4 \, L^3} \, e^{2 \pi i \vec{n} \langle \vec{X}
\rangle /L}.
\end{equation}
Here the energy in the center of mass frame of the gauge bosons is
denoted by $\smash{\sqrt{s}}$. Let $\smash{\langle \vec{X_1} \rangle
}$ and $\smash{\langle \vec{X_2} \rangle }$ be the positions of the
two brane stacks inside the $T^6$.  If we denote the relative distance
of the stacks by $\smash{\vec{A} \equiv \langle \vec{X_2} \rangle -
\langle \vec{X_1} \rangle }$ and introduce the shorthand $\vec{a}
\equiv 2 \pi \vec{A} /L$, the matrix element corresponding to the
process in Fig.~\ref{scat} is given by
\begin{equation}
\label{Matrixelement}
\mathcal{M} \, \sim \, \frac{s^2}{ M_{10}^8 \, L^6} \, \sum_{\vec{n}
\in \mathbb{Z}^6 } \, \frac{ e^{ i \vec{n} \vec{a}}}{s -m_{\vec{n}}^2
+i \epsilon}.
\end{equation}
We have ignored prefactors of order one. For phenomenological
purposes, we can safely assume $\smash{\sqrt{s} < L^{-1}}$. Namely,
since the energy $\smash{\sqrt{s}}$ of the colliding gauge bosons is
determined by the temperature $T$ of the heated gauge theory, this
corresponds to $\smash{T < L^{-1}}$. If this were not the case, the
gauge theory would heat up the compact manifold and the geometrical
picture would be lost.  Following from $\smash{\sqrt{s} < L^{-1}}$,
one has $s < m_n^2$ for $n >0$ and the contribution of the energy
$\smash{\sqrt{s}}$ in the propagator can be neglected for all but the
zero mode. Thus, Eq.~\eqref{Matrixelement} simplifies to
\begin{equation}
\label{Matrixelement2}
\mathcal{M} \, \sim \, \frac{s^2}{ M_{10}^8 \, L^4} \, \sideset{}{'}
\sum_{\vec{n} \in \mathbb{Z}^6 } \, \frac{ e^{ i \vec{n} \vec{a}}}
{\vec{n}^2} + \frac{s}{ M_{10}^8 \, L^6},
\end{equation}
where the prime denotes exclusion of $\smash{\vec{n} = \vec{0}}$ in
the sum. Since the 4d Planck scale is determined by $M_4^2\simeq
M_{10}^8 L^6$, the last term in Eq.~\eqref{Matrixelement2} simply
reflects the fact that the zero mode interacts with gravitational
strength. The sum, which would be UV divergent in absence of the
exponential factor, is dominated by terms with large
$\smash{\vec{n}}$.  It can therefore be approximated by an integral:
\begin{equation}
\label{sum}
\int d^6 n \, \frac{ e^{ i \vec{n} \vec{a}}}{\vec{n}^2} \sim
\frac{1}{a^4}.
\end{equation}
The r.h. side of Eq.~\eqref{sum} results from the fact that the
exponential function oscillates quickly for $|\smash{\vec{n}| \gtrsim
a^{-1} }$ ($\smash{a \equiv |\vec{a}|}$), effectively cutting off the
integral.\footnote{ One can see in particular that the sum in
Eq.~\eqref{Matrixelement2} is effectively cut off before the geometry
of the throats becomes relevant, justifying our flat-space
approximation.  } More precisely, we evaluate a similar but more
general integral, which we will need in
Sect.~\ref{decayingaugepicture2}, in the Appendix.
Equation~\eqref{sum} follows from this integral in a particular limit,
which is displayed in Eq.~\eqref{lc}.

Inserting Eq.~\eqref{sum} into Eq.~\eqref{Matrixelement2}, we find
\begin{equation}
\label{Matrixelement3}
\mathcal{M} \, \sim \, \frac{s^2}{ M_{10}^8 \, A^4} + \frac{s}{
M_{10}^8 \, L^6}\,,
\end{equation}
where $\smash{A \equiv |\vec{A}|}$. For an order-of-magnitude
calculation, we can neglect the interference term in
$\smash{|\mathcal{M}|^2}$. The cross section for the process in
Fig.~\ref{scat} then reads
\begin{equation}
\label{sigma}
\sigma \, \sim \, \frac{s^3}{M_{10}^{16} A^8} + \frac{s}{M_{10}^{16}
L^{12}} \qquad\text{for}\qquad\sqrt{s} < L^{-1}\,.
\end{equation}
Inserting this cross section into Eq.~\eqref{rate}, we get the energy
loss rate due to scattering of one particle species into another
particle species.  To get the total energy loss rate, we have to sum
over all initial and final state species and polarizations. Let us
denote with $N_1$ and $N_2$ the number of colors of the heated gauge
theory and the gauge theory that is being heated, respectively. The
summation then gives extra factors of $2N_1^2$ and $2N_2^2$ and we
get, again neglecting prefactors of order one coming from the
integration in Eq.~\eqref{rate},
\begin{equation}
\dot{\rho} \sim \frac{N_1^2 N_2^2}{M_{10}^{16} A^8} \, T^{13} +
\frac{N_1^2 N_2^2}{M_{10}^{16} L^{12}} \, T^9.
\end{equation}
Using Eq.~\eqref{RN}, this can be written in a slightly more compact
form. Denoting by $R_1$ and $R_2$ the AdS scales of the corresponding
throats, we arrive at the main result of this section:
\begin{equation}
\label{el}
\dot{\rho} \sim \frac{R_1^8 R_2^8}{A^8} \, T^{13} + \frac{R_1^8 R_2^8}
{L^{12}} \, T^9.
\end{equation}

An apparent limitation of our analysis is the assumption of a simple
toroidal geometry for the embedding space. This assumption was used to
determine the spectrum and the couplings of higher KK modes (which
determine the first term in Eqs.~\eqref{Matrixelement2} and
\eqref{el}). By contrast, the coupling of the zero mode (which
determines the second term in Eqs.~\eqref{Matrixelement2}
and~\eqref{el}), depends only on the size of the embedding manifold
and not on its geometry. To see the relative importance of the terms
more clearly, we rewrite Eq.~\eqref{el} as
\begin{equation}
\label{el2}
\dot{\rho} \sim \frac{R_1^8 R_2^8}{A^8} \, T^{13} \left( 1 + \left(
\frac{A}{L} \right)^8 \left( L \, T \right)^{-4} \right).
\end{equation}
If the throat-to-throat distance is large, $A\sim L$, the second term
dominates (recall that $LT<1$) and the precise geometry is
irrelevant. By contrast, for small throat separation, $\smash{A \ll
L(LT)^{1/2}}$, the contribution of the KK modes is dominant. In this
case, the precise geometry of the embedding manifold may in principle
be relevant. However, it is then natural to assume that the curvature
scale in the region between the throats is smaller than
$1/A$. Furthermore, as we have already pointed out above, the sum in
Eq.~\eqref{Matrixelement2} is dominated by contributions with
$|\vec{n}|\sim L/A$, corresponding to masses $\smash{m_{\vec{n}} \sim
A^{-1}}$. Such modes are only sensitive to the geometry at distance
scales $A$ in the vicinity of the two throats, which we just argued to
be approximately flat. Thus, the order of magnitude of our result will
remain correct in most relevant cases, even if the overall geometry is
very different from that of a torus.

In particular, we see that O-planes and further D-brane stacks will
not change our result as long as they are not too close to the two
throats. Moreover, we can apply our result to situations with one
Klebanov-Strassler throat and one AdS$_5 \times $S$^5$ throat or with
two Klebanov-Strassler throats as long as the curvature scale of the
space in between the two throats is not much larger than $1/A$.

In order for the calculation in terms of gauge fields to be justified,
the temperature of the heated throat has to be larger than its
IR/confinement scale.\footnote{ Otherwise, the heated throat sector
contains a non-relativistic gas of KK modes, whose decay rate to the
other throat will be determined in Sect.~\ref{KKdecay}.  } One can
then easily see from the gravity picture that the finite length of the
Klebanov-Strassler throats will not change the result qualitatively.
This is obvious for the heated throat since the black hole horizon
hides the IR region. For the throat to which the energy is
transferred, the argument is as follows: In the gravity picture,
energy transfer is due to Hawking radiation, which is emitted by the
heated throat and subsequently absorbed by the other throat. But only
the geometry in the UV region of the throat is important for the
absorption by (or, equivalently, the tunneling into) that throat.

\section{Decay of KK modes between two throats}\label{KKdecay}
Another interesting quantity for cosmology is the rate with which KK
modes localized in one throat decay to a different throat. This
question has already received significant attention in the literature
(see~\cite{
smallnumbers,Barnaby:2004gg,Kofman:2005yz,Frey:2005jk,Chialva:2005zy,CT,
Firouzjahi:2005qs,Langfelder}), mainly in the context of reheating
after brane-antibrane inflation. However, in all cases the
calculations were done in the gravity picture, whereas we will again
(mainly) exploit the gauge theory point of view. This will allow us to
incorporate easily the dependence on the throat radii and the distance
between the throats. We compare the results from the literature with
ours in Sect.~\ref{comparison}.

\subsection{The glueball decay vertex}\label{decayingaugepicture}
\begin{figure}[t]
\begin{minipage}{6.2cm}
\begin{center}
\vspace{10mm} \includegraphics[scale=0.6]{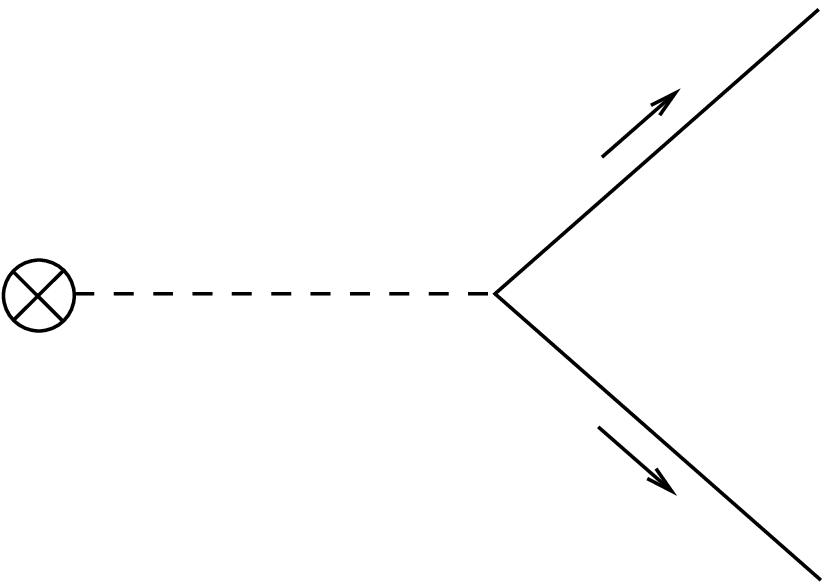}
\vspace{6.5mm}
\caption{Feynman diagram for the decay of a glueball in one throat
into fields in another throat.\label{dec}}
\end{center}
\end{minipage}
\hspace{0.8cm}  \begin{minipage}{8.2cm}
\vspace{-12mm}
\begin{center}
\includegraphics[scale=0.8]{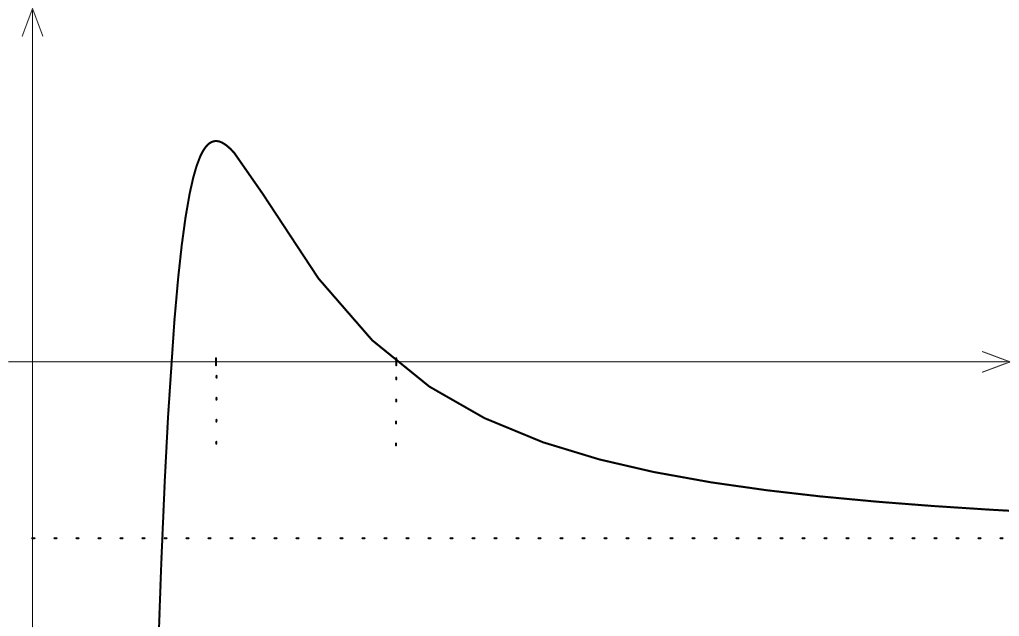}  \put(-240,160){$V/m^2$}  
\put(-245,28){$-1$}  \put(-195,38){\footnotesize$\sim mR$}  
\put(-155,38){\footnotesize$\sim (mR)^{-1}$}  \put(-50,55){$R/r=z/R$}
\caption{Potential in the effective Schr\"o\-din\-ger equation for the
dilaton in a throat.\label{pot}}
\end{center}
\end{minipage}
\end{figure}

We want to calculate the decay rate of glueballs on one brane stack
into two gauge fields on another brane stack. As in
Sect.~\ref{energytransfer}, we perform the calculation for two
D3-brane stacks in a 6-dimensional torus of uniform size $L$. As
before, we can argue that our result provides the right order of
magnitude also for more general geometries. The Feynman diagram for
the process is shown in Fig.~\ref{dec}. Due to the non-renormalization
theorem described in the introduction, we do not have to care whether
the decay products will arrange into one or more glueballs. The vertex
for this part of the diagram is simply the one already derived in
Eq.~\eqref{vertex}. However, the other vertex between a dilaton and a
glueball can not so easily be read off from the Lagrangian. Therefore,
we make use of the gravity picture to calculate the decay rate in a
simpler situation. From this we will determine the vertex by demanding
that this decay rate agree with the gauge theory picture.

Namely, we consider a dilaton localized in a single AdS$_5
\times$S$^5$ throat which is embedded into flat 10d space. This is the
geometry of an extremal black 3-brane, the metric being given
by~\cite{Horowitz:1991cd}
\begin{eqnarray}
\label{threebrane}
ds^2 &=& f(r)^{-1/2} \, \left( -dt^2+dx_1^2+dx_2^2+dx_3^2 \right) +
f(r)^{1/2} \, \left( dr^2 + r^2 d\Omega_5^2 \right)\nonumber \\
&&\text{with}\qquad f(r)= 1+ \frac{R^4}{r^4}.
\end{eqnarray}
In the near-horizon region, for $\smash{r \ll R}$, the warp factor
reduces to $f(r)\simeq R^4/r^4$ and the geometry is asymptotically
AdS$_5\times$S$^5$.  Far from the horizon, for $r \gg R$, the warp
factor is $f(r) \simeq 1$ and the geometry is asymptotically flat 10d
space. In this situation, the AdS/CFT conjecture is based on taking
the near-horizon limit $r \rightarrow 0$ and $\alpha' \rightarrow 0$,
while keeping $r/\alpha'$ fixed.  This effectively reduces the
geometry to the AdS$_5 \times$S$^5$ part. In the equivalent
description by a stack of D3-branes in flat space, interactions
between supergravity and the world-volume theory vanish in this
limit. This can be seen, e.g., from Eq.~\eqref{coupling2}, since
$\alpha' \rightarrow 0$ implies $\kappa_{10} \sim g_s \alpha'^2
\rightarrow 0$ at finite $g_s$.  One then identifies states in the
world-volume gauge theory with eigenmodes of supergravity on AdS$_5
\times$S$^5$.

In our considerations, however, we want to retain the asymptotically
flat part of the geometry. What were previously the eigenmodes on
AdS$_5 \times$S$^5$ will then become part of the spectrum of
excitations in the full geometry, including excitations outside of the
throat region.  This reflects the fact that, since we do not set
$\alpha'$ to zero, the gauge theory will interact with the
supergravity fields in the embedding space.

Such nonvanishing interactions lead to the decay of gauge theory
states, to which we refer as glueballs, into supergravity fields. This
glueball decay has a simple counterpart on the gravity side. Namely,
excitations in the gauge theory correspond to excitations in the
throat region. The state dual to the glueball will therefore be a wave
packet which is localized in the throat.  Due to the different time
evolution of its constituent modes, this wave packet will decohere
after a certain time (see \cite{Langfelder}). Hence, excitations will
show up in the asymptotically flat region as well, which is the
analogue of glueball decay.

We will now determine the decay rate of a dilaton localized in the
throat into flat 10d space. To this end, we will assume the throat to
be sharply cut off somewhere in the IR. Such an AdS$_5 \times$S$^5$
throat with an IR cutoff might not exist as a solution to
supergravity, but it can serve as a simple toy model capturing the
relevant information.  Later on we will show how to extend our results
to realistic finite throats such as the Klebanov-Strassler
throat~\cite{KS}. On the gauge theory side, the cutoff corresponds to
a deformation by a relevant operator, in which case the gauge theory
has a discrete set of glueball states.

The wave equation for the dilaton is just the Laplace equation in the
background geometry:
\begin{equation}
\label{LE}
\partial_M \left( \sqrt{g} g^{MN} \partial_N \phi \right) = 0.
\end{equation}
Using Eq.~\eqref{threebrane}, one gets
\begin{equation}
\label{eom}
\left[ r^{-5} \frac{d}{dr} \, r^5 \, \frac{d}{dr} + m^2 + \frac{m^2
R^4} {r^4} - \frac{l(l+4)}{r^2} \right] \phi(r) = 0,
\end{equation}
where $l (l+4)$ is the eigenvalue of the Laplacian on S$^5$ and $m^2$
is the eigenvalue of the 4d d'Alembertian. We will call $m$ the mass
of the excitation. Choosing a new radial coordinate $\smash{z \equiv
R^2 /r }$ and introducing a redefined field $\smash{\tilde{\phi}
\equiv z^{-3/2} \phi}$, we arrive at
\vspace{1mm}
\begin{equation}
\label{se}
\frac{d^2}{dz^2} \, \tilde{\phi}(z) + \left( m^2 - \frac{15/4 + l
(l+4)} {z ^2} + \frac{m^2 R^4}{z^4} \right) \, \tilde{\phi}(z) = 0.
\end{equation}

This has the form of a Schr\"odinger equation, the potential being
given by the term in brackets. A schematic plot of this potential is
shown in Fig.~\ref{pot}. As one can see, a wave coming from the
near-horizon region ($z \rightarrow \infty $ corresponding to $r
\rightarrow 0$) has to tunnel through an effective barrier to reach
the asymptotically flat region ($z \rightarrow 0$ corresponding to $r
\rightarrow \infty$).\footnote{ As we will see in
Sect.~\ref{comparison}, by using cartesian coordinates for the torus,
one again gets a Schr\"odinger-like equation. However, in this case
there is no barrier an incoming wave would have to tunnel
through. Instead, the reflection of a large part of the incoming wave
is due to the steepness of the potential well.  } The tunneling
probability $\mathcal{P}$ has been calculated in~\cite{K} (see
also~\cite{CT}) for masses $\smash{m \ll R^{-1}}$,
\begin{equation}
\label{tp}
\mathcal{P} \sim \, \left(m R \right)^{8+4 l}\,,
\end{equation}
where we again neglect prefactors of order one.

Although this result has been derived for a throat which is infinite
in the IR direction, we can still use it for a finite throat, as long
as the mass $m$ of the wave is not too small. For a throat which is
cut off in the IR at $z = z_\IR$, masses are quantized in units of
$m_\IR \equiv \smash{z_\IR^{-1}}$ such that $m_n \sim n \, m_\IR$ with
$n$ an integer.\footnote{ Solutions to Eq.~\eqref{se} in the throat
region ($z \gg R$) are $\smash{\tilde{\phi} \simeq A \, \sqrt{m z}
J_{l+2}(m z) + B \, \sqrt{m z} Y_{l+2}(m z)}$, where $A$ and $B$
follow from the boundary condition on the UV side of the throat and
from normalization. For sufficiently large $z$, $\smash{\tilde{\phi}}$
behaves as $\smash{\tilde{\phi} \sim A \cos m z+ B \sin m z}$.  The
quantization of $m$ is a result of the boundary condition for
$\tilde{\phi}$ or its first derivative at $z=z_\IR$.  } The result of
Eq.~\eqref{tp} can then be trusted as long as the wave function is not
completely dominated by the unknown IR cutoff region. This will be the
case if $n$ is sufficiently larger than 1.

The wave packet describing the glueball can be decomposed into a set
of modes moving in the IR direction and in the UV direction. If the
barrier on the UV side were impenetrable, the modes would be reflected
entirely on the UV and IR side. However, since a small fraction of the
incoming flux is able to penetrate the barrier, the wave leaks out of
the throat. The incoming and outgoing fluxes at the barrier,
$j_{\text{in}}$ and $j_{\text{out}}$, determine the tunneling
probability $\mathcal{P}$ and the decay rate $\Gamma$:
\begin{equation}
\label{dr}
\mathcal{P}=1-j_{\text{out}}/j_{\text{in}}\quad,\quad\Gamma =
j_{\text{in}} \mathcal{P}.
\end{equation}
Thus, a wave packet localized in the throat will decohere.

To determine $\Gamma$, we need solutions to Eq.~\eqref{se} describing
waves which are reflected back and forth between the UV barrier and
the IR end of the throat. From these we can calculate the incoming
flux $j_{\text{in}}$.  We restrict ourselves to the case $\smash{m \ll
R^{-1}}$. In particular, this means that $z_\IR \gg R$, where $z= R$
corresponds to the beginning of the throat region
(cf.~Eq.~\eqref{threebrane}). For $\smash{z \gg m^{-1} \gg R}$, we can
neglect the last two terms in the potential, keeping only the constant
term $m^2$. In this limit, the solution is simply given by plane
waves:
\begin{equation}
\label{planewave}
\tilde{\phi} \simeq A \, \cos{m z} + B \, \sin{m z}.
\end{equation}
The approximation is valid for $\smash{z_\IR \geq z \gg m^{-1}} \sim
z_\IR/n$.  If $n$ is not too small, the mode is well approximated by a
plane wave in a large portion of the throat. Deviations from this form
for $z \lesssim z_\IR/n$ are due to reflection at and tunneling
through the effective barrier.

To calculate $j_{\text{in}}$ from Eq.~\eqref{planewave}, we have to
determine the normalization of the solution in physical terms. As a
simplification, we consider a complex scalar and a plane wave moving
around an $S^1$ parametrized by $z \in [0,z_\IR)$. Going to the rest
frame with respect to momenta parallel to the brane and reinstating
time dependence, we have
\begin{equation}
\label{testwave}
\tilde{\phi} = \mathcal{N} \, e^{i m (z + t)}
\end{equation}
for the plane wave moving towards the UV barrier. To determine the
normalization constant $\mathcal{N}$, we use the standard charge
density for a Klein-Gordon particle, $\smash{j^0=\Im(\tilde{\phi}^*
\partial_t \, \tilde{\phi})}$. It has to be normalized according to
\begin{equation}
\label{normalization}
1 = \int_0^{z_\IR} dz \, j^0 \; \; \Rightarrow \; \; \mathcal{N} =
\frac{1} {\sqrt{m z_\IR}}.
\end{equation}
The flux is then given by $j_{\text{in}}=j^z = \Im(\tilde{\phi}^*
\partial_z \, \tilde{\phi})$. Using the solution of
Eq.~\eqref{testwave} with the normalization of
Eq.~\eqref{normalization}, we find
\begin{equation}
\label{flux}
j_{\text{in}} = \frac{1}{z_\IR}=m_\IR.
\end{equation}
Using this result and Eq.~\eqref{tp}, the decay rate of a glueball
follows from Eq.~\eqref{dr} as
\begin{equation}
\label{Gamma}
\Gamma \sim m_\IR (m R)^{8+4 l}.
\end{equation}

Now that we have the decay rate in the gravity picture, we need to
define a vertex $V$ in the gauge theory picture which reproduces this
result. We model the coupling by a term 
\be 
{\cal L_{\rm
10d}}\,\,\supset\,\, V\,\delta^{(6)}(\vec{X}-\langle\vec{X}
\rangle)\,\phi(x,\langle\vec{X}\rangle)\,{\cal G}(x) 
\ee 
in the 10d
Lagrangian, where $\mathcal{G}$ denotes the glueball state with
canonically normalized 4d kinetic term. Compactifying the 6 dimensions
perpendicular to the brane on a torus of size $L$ for the moment and
using the KK mode decomposition of Eq.~\eqref{expansion}, we get the
effective 4d Lagrangian
\begin{equation}
\label{effectiveL}
\mathcal{L}_{\rm 4d} \, \supset \, \sum_{\vec{n} \in \mathbb{Z}^6}
\left( - \frac{1}{2}\partial_{\mu} \Phi_{\vec{n}} \, \partial^{\mu}
\Phi_{\vec{n}} - \frac{1}{2} m_{\vec{n}}^2 \, \Phi_{\vec{n}}^2 + e^{2
\pi i \vec{n} \langle \vec{X} \rangle /L} \,\frac{V}{L^3} \,
\Phi_{\vec{n}}(x) \, \mathcal{G}(x) \right)\,.
\end{equation}
From this, the total decay rate of a glueball into KK modes of the
dilaton follows:
\begin{equation}
\Gamma = \frac{1}{2 \omega_i} \frac{1}{L^6} \sum_{\vec{n} \in
\mathbb{Z}^6} \int \frac{d^3 p_f}{(2 \pi )^3} \,\frac{1}{2 \omega_f}
\, (2 \pi)^4 \, \delta^{(4)}(p_f -p_i) \, |V|^2.
\end{equation}
In this formula, $p_f=p_{f_\parallel}$ is a 4-vector characterizing
the momentum of the final-state dilaton parallel to the brane, while
$\omega_i$ and $\omega_f$ are the energies of the initial and final
state.  The 4-momentum of the decaying glueball is denoted by
$p_i$. Introducing the dilaton momentum in the compact dimensions as
$\smash{\vec{p}_{f_\perp} = 2 \pi \,\vec{n} /L}$, we can replace the
sum by an integral when we go back to $L \rightarrow \infty$:
\begin{equation}
\frac{1}{L^6} \, \sum_{\vec{n} \in \mathbb{Z}^6} \; \longrightarrow \;
\int \frac{d^6p_{f_\perp}}{(2 \pi)^6}\,.
\end{equation}
The decay rate of a glueball into a dilaton is then given by
\begin{equation}
\label{dr2}
\Gamma = \frac{1}{2 \omega_i} \int \frac{d^6 p_{f_\perp}}{(2 \pi )^6}
\frac{d^3 p_{f_\parallel}}{(2 \pi )^3} \,\frac{1}{2 \omega_f} \, (2
\pi)^4 \, \delta^{(4)}(p_{f_\parallel }-p_i) \, |V|^2\,.
\end{equation}
Since the dilaton is massless, $\smash{\omega_f=
  \sqrt{|\vec{p}_{f_\perp}|^2+ |\vec{p}_{f_\parallel}|^2}}$. Going to
the rest frame of the glueball, $\smash{\vec{p_i}} = 0$, and
performing the momentum integrations, we arrive at
\begin{equation}
\Gamma = \frac{1}{2 \omega_i} \int \frac{d^6 p_{f_\perp}}{(2 \pi )^6}
\,\frac{1}{2 \omega_f} \, (2 \pi) \, \delta(\omega_f - \omega_i) \,
|V|^2 \sim \omega_i^3 \, |V|^2\,,
\end{equation}
where we have used $\smash{\omega_f=|\vec{p}_{f_\perp}|}$ and
neglected prefactors of order one. In its rest frame, $\omega_i$ is
simply the mass $m$ of the glueball. Comparing with Eq.~\eqref{Gamma},
we get
\begin{equation}
\label{vertex2}
V \sim \sqrt{m_\IR m}\,\,\,m^{2+ 2l}R^{4+2 l}.
\end{equation}

\subsection{Decay rate calculation in the gauge theory picture}
\label{decayingaugepicture2}
With the effective vertex $V$ at hand, calculating the decay rate of
one glueball into gauge fields living on a different brane stack is
straightforward. Following from Eqs.~\eqref{effectiveL} and
\eqref{vertex2}, the vertex between a glueball and a KK mode of the
dilaton is
\begin{equation}
\frac{V}{L^3} \, e^{2 \pi i \vec{n} \langle \vec{X} \rangle /L}.
\end{equation}
The other vertex in the diagram is still given by
Eq.~\eqref{vertex}. Summing over all intermediate KK modes, we arrive
at an expression very similar to Eq.~\eqref{Matrixelement}:
\begin{equation}
\label{Matrixelement5}
\mathcal{M} \, \sim \, \frac{\sqrt{m_\IR m} \, ( m R )^{4+ 2 l}}{
M_{10}^4 \, L^6} \, \sum_{\vec{n} \in \mathbb{Z}^6 } \, \frac{ e^{ i
\vec{n} \vec{a}}}{m^2 -m_{\vec{n}}^2 +i \epsilon}.
\end{equation}
Compared to Eq.~\eqref{Matrixelement}, the only difference is the
prefactor and the substitution of the energy $\smash{\sqrt{s}}$ of the
colliding gauge bosons by the mass $m$ of the glueball.

We will analyse Eq.~\eqref{Matrixelement5} in two different regimes,
namely for $m^{-1}>L$ and for $m^{-1}\ll~L$. The former case is the most
interesting one from a phenomenological viewpoint. As we argued in
Sect.~\ref{energytransfer}, we can assume that the reheating
temperature $T_{\text{RH}}$ in early cosmology is smaller than
$L^{-1}$. Accordingly, the mass $m$ of any relic KK modes is also
restricted by $m<L^{-1}$. The latter case, on the other hand, can be
easily analysed in the gravity picture as well. We will perform this
cross-check in Sect.~\ref{comparison}.

For $m^{-1}>L$, we can make the same simplifications as in
Eq.~\eqref{Matrixelement2} and use Eq.~\eqref{sum} for the sum. The
decay rate of a glueball into a pair of gauge bosons follows from the
standard 4d formula: 
\be 
\Gamma\sim m^{-1}|{\cal M}|^2\,.\label{4dg}
\ee 
To get the total decay rate, we have to sum over the $N^2$ final
state gauge bosons. If we denote by $R_1$ and $R_2$ the AdS scale of
the throat containing the initial and the final state, respectively,
we find
\begin{equation}
\label{DecayRate}
\Gamma \sim \frac{R_1^{8+4 l} R_2^8}{ A^8} \, m_\IR \, m^{8+4 l}+
\frac{R_1^{8+4 l} R_2^8}{L^{12}} \, m_\IR \, m^{4 + 4 l}.
\end{equation}
Although the derivation of this decay rate assumed two AdS$_5
\times$S$^5$ throats and a torus as embedding manifold, it can also be
applied to more general geometries, according to the discussion in
Sect.~\ref{energytransfer}.  However, for different throat geometries
the dependence on the eigenvalues of the angular Laplacian is of
course different. These eigenvalues entered the discussion through the
tunneling probability Eq.~\eqref{tp}, from which we determined the
dilaton-glueball vertex in Eq.~\eqref{vertex2}. For the example of a
Klebanov-Strassler throat~\cite{KS}, let us outline how to determine
the dilaton-glueball vertex for more general throat geometries. Away
from the bottom of the throat at $r=r_s$, the warp factor of a
Klebanov-Strassler throat is well approximated by
\begin{equation}
\label{KSwf}
A(r) = 1 +\frac{R^4 \, \ln(r/r_s)}{r^4}.
\end{equation}
The effective AdS scale $R$ depends on the number of fractional
D3-branes at the conifold singularity. The metric is still given by
Eq.~\eqref{threebrane} away from $r = r_s$ if one also replaces the
line element $d\Omega_5^2$ of a sphere by the line element of
$T^{1,1}$. For $R\gg r \gg r_s$, which defines the throat region, the
warp factor is approximately $A \simeq R^4 \ln(r/r_s)/r^4$. For $r\gg
R$, where the geometry is asymptotically a cone over $T^{1,1}$, we
have $A \simeq 1$. Near $r=r_s$, the geometry differs considerably
from Eqs.~\eqref{KSwf} and~\eqref{threebrane} and the throat is cut
off by the Klebanov-Strassler region. For an order of magnitude
estimate, one can neglect the logarithmic $r$ dependence of the warp
factor away from $r=r_s$ and approximate the Klebanov-Strassler region
by a sharp cut off~\cite{BHT}. Thus, the tunneling probability from
the throat into the conical region can be (approximately) calculated
from the effective Schr\"odinger equation, Eq.~\eqref{se}. The
dependence on the eigenvalues of the Laplacian on $T^{1,1}$ enters
through the potential, where they replace the corresponding
eigenvalues $l (l+4)$ on an $S^5$. Moreover, for an AdS warp factor
and a sharp cut off, the incoming flux is given by Eq.~\eqref{flux},
as before. From Eq.~\eqref{dr}, one can then determine the decay rate
and match the vertex such that this decay rate is reproduced.

Let us now consider the case $m^{-1}\ll L$. We will also assume $A\ll
L$ for simplicity. Recalling that $\smash{m_{\vec{n}}=2 \pi
|\vec{n}|/L}$ and $\vec{a} = 2 \pi \vec{A} /L$, we can approximate the
sum in Eq.~\eqref{Matrixelement5} by an integral,
\begin{equation}
\label{propagator}
\frac{1}{L^6} \, \sum_{\vec{n} \in \mathbb{Z}^6 } \, \frac{ e^{2 \pi i
\vec{A} \, \vec{n}/L}}{m^2 -(2 \pi)^2 \vec{n}^2/L^2 +i \epsilon}
\;\;\; \longrightarrow \;\;\; \int \frac{d^6 \rho}{(2 \pi)^6} \,
\frac{ e^{i \vec{A} \, \vec{\rho}}}{m^2 -\vec{\rho}^2 +i \epsilon},
\end{equation}
where $\vec{\rho} \equiv 2 \pi \,\vec{n} /L$. The resulting expression
is just the propagator of a massless particle in a mixed,
energy-configuration-space representation, with the `energy' $m$
characterizing the invariant 4-momentum. This is of course expected in
the large $L$ limit, where the torus goes over to flat space and the
infinite KK tower is replaced by the underlying higher-dimensional
dilaton field. The integral is evaluated in the Appendix, the outcome
being
\begin{equation}
\label{HankelFunktion}
\int \frac{d^6 \rho}{(2 \pi)^6} \, \frac{ e^{ i \vec{A} \,
\vec{\rho}}}{m^2 -\vec{\rho}^2 +i \epsilon} \sim \frac{m^2}{A^2} \,
H_2^+( m A),
\end{equation}
where $H_2^+(x)=J_2(x)+i\, Y_2(x)$ is a Hankel function and we have
neglected prefactors of order one. Using the asymptotic forms of the
Bessel functions for large and small arguments,
Eq.~\eqref{HankelFunktion} can be simplified as follows:
\begin{equation}
\label{lc}
\frac{m^2}{A^2} \, H_2^+( m A) \sim
\begin{cases} 
\frac{m^{3/2}}{A^{5/2}} \, e^{i \, m A} & \text{for } m^{-1}\ll A \\
\frac{1}{A^4} & \text{for } m^{-1}\gg A\,.
\end{cases}
\end{equation}
Inserting these results in Eq.~\eqref{Matrixelement5}, we get the
matrix elements $\mathcal{M}$ for these two cases. The corresponding
partial decay rates follow from Eq.~\eqref{4dg}. Summing over all
final state species, we find
\begin{equation}
\label{DecayRate2}
\Gamma \sim
\begin{cases} 
\frac{R_1^{8+4 l} R_2^8}{A^{5}} \, m_\IR \, m^{11+4 l} & \text{for }
m^{-1}\ll A \\ \frac{R_1^{8+4 l} R_2^8}{A^8} \, m_\IR \, m^{8+4 l} &
\text{for } m^{-1}\gg A
\end{cases}.
\end{equation}
Again, the same discussion as before applies concerning the extension
to more realistic geometries. As a consistency check, we should
examine, whether the appropriate limiting cases of
Eqs.~\eqref{DecayRate} and~\eqref{DecayRate2} coincide. The regions of
validity of the two calculations have a common border for $A\ll
m^{-1}\sim L$. Indeed, for this choice of parameters the first term in
Eq.~\eqref{DecayRate} dominates and the result agrees with the second
line of Eq.~\eqref{DecayRate2}.

\subsection{Some calculations in the gravity picture}
\label{comparison}
As in the sections before, we consider two AdS$_5 \times$S$^5$ throats
embedded in a 6-dimensional torus of uniform size $L$. The geometry is
that of a multi-centered black 3-brane, the metric being
\begin{equation}
\label{multicenter}
ds^2 = A^{-1/2} \, \left( -dt^2+dx_1^2+dx_2^2+dx_3^2 \right) + A^{1/2}
\, \left( dx_4^2 + \cdots + dx_9^2 \right)
\end{equation}
with
\begin{equation}
A(\vec{x}_{\perp}) = 1 + \sum_{\vec{n} \in \mathbb{Z}^6 } \left(
\frac{R_1^4}{|\vec{x}_{\perp} -\vec{A}_1 + \vec{n} L|^4} +
\frac{R_2^4}{|\vec{x}_{\perp} -\vec{A}_2 + \vec{n} L|^4} \right).
\end{equation}
The positions of the two throats are denoted by $\vec{A}_1$ and
$\vec{A}_2$, their AdS scales by $R_1$ and $R_2$. The vector
$\smash{\vec{x}_\perp}$ refers to the coordinates $x_4,\dots, x_9$ in
the torus. The sum in the warp factor $\smash{A(\vec{x}_{\perp})}$ is
due to mirror effects in the torus. Again, this is not a consistent
compactfication. Including O-planes, for example, would give extra
contributions to the warp factor (see~\cite{Verlinde}). We try to
calculate the transition of a dilaton between different throat
regions, which is the gravity counterpart to the gauge theory
calculation in Sects.~\ref{decayingaugepicture}
and~\ref{decayingaugepicture2}. The equation of motion for the dilaton
is given in Eq.~\eqref{LE}. Inserting Eq.~\eqref{multicenter} in
Eq.~\eqref{LE} and using $\smash{\sqrt{g}=A^{1/2}}$, one gets
\begin{equation}
\partial_n \, \partial^n \, \phi + A(\vec{x}_{\perp}) \,
\partial_{\mu} \, \partial^{\mu} \, \phi = 0.
\end{equation}
The indices $\mu$ and $n$ run from 0 to 3 and from 4 to 9,
respectively. Using the 4d Klein-Gordon equation, one arrives at
\begin{equation}
\label{se2}
\partial_n \, \partial^n \, \phi + A(\vec{x}_{\perp}) \, \, m^2 \,
\phi = 0,
\end{equation}
where $m$ is the kinetic energy perpendicular to the branes. Like
Eq.~\eqref{se}, this has the form of a Schr\"odinger
equation. Contrary to Eq.~\eqref{se}, however, there is no potential
barrier separating the throat region and asymptotically flat space,
since the potential $\smash{V = - m^2 A(\vec{x}_{\perp})}$ is strictly
negative. The difference comes from using cartesian coordinates
perpendicular to the branes in Eq.~\eqref{multicenter} rather than
spherical coordinates in Eq.~\eqref{threebrane}. Still, a wave in the
throat region, moving away from the horizon, is reflected to a large
part before entering asymptotically flat space. In cartesian
coordinates, however, this is due to the steepness of the potential
well.

To determine the transition probability $\mathcal{P}$ of a dilaton
between two throat regions, one has to solve Eq.~\eqref{se2} with
appropriate boundary conditions. Then $\mathcal{P}$ is the ratio of
incoming flux in one throat and outgoing flux in the other throat. In
general, the corresponding calculation is difficult. However, if the
torus is very large ($L \rightarrow \infty$) and the throats are
sufficiently far apart ($A \gg \smash{m^{-1}}$), the problem
effectively splits into two simpler calculations. Namely, the latter
condition means that the de Broglie wavelength of the particle is
small compared to the distance of the throats. A transition between
two throats can then be described as a two-step process. For
simplicity, we take the initial state in the first throat to be an
s-wave. Only a small fraction of the outgoing flux reaches the
asymptotically flat region, the probability being (cf. Eq.~\eqref{tp}
for $l=0$)
\begin{equation}
\mathcal{P}_1 \sim \left(m R_1 \right)^8.
\end{equation}
In between the two throats, one has a free spherical wave,
approximating a plane wave near the second throat. The absorption
cross section (per brane world-volume) for such a plane wave was
calculated in~\cite{K}. Neglecting prefactors of order one, it reads
\begin{equation}
\sigma_2 \sim m^3 R_2^8.
\end{equation}
Near the second throat, the incoming flux will be diluted by a factor
of $\smash{A^{-5}}$, since the free spherical wave is expanding in
6-dimensional flat space. The absorption probability by the second
throat thus is
\begin{equation}
\mathcal{P}_2 \sim \frac{\sigma_2}{A^5} \sim \frac{m^3 R_2^8}{A^5}.
\end{equation}
The transition probability between the two throats is just the product
$\mathcal{P}_1 \mathcal{P}_2$. If we denote by $m_\IR$ the mass gap in
the first throat, using Eqs.~\eqref{dr} and~\eqref{flux} the decay
rate from the gravity calculation follows as
\begin{equation}
\Gamma \sim \frac{R_1^8 R_2^8}{A^5} \, m^{11} m_\IR.
\end{equation}
This is precisely what we found in Eq.~\eqref{DecayRate2} for $A \gg
m^{-1}$ and $l=0$.  The crucial ingredient is the $A^{-5}$
dependence. That it agrees in both calculations is, however, not too
surprising. In the gauge theory calculation, it came from the
propagator in a mixed energy-configuration-space representation
(cf. Eq.~\eqref{propagator}). The same is of course true in the above
gravity calculation, although we have not stated it explicitly.

There is yet another situation where the decay rate between two
throats is comparatively easy to obtain. Let us consider only one
throat for the moment. We do not need to specify the precise form, but
will assume that it is finite and reasonably well approximated by a
slice of AdS$_5$ times some compact manifold $\mathcal{M}$. The prime
example certainly is a Klebanov-Strassler throat, whose interpretation
as a stabilized Randall-Sundrum model was given in~\cite{BHT}. Let us
denote by $R_1$ the (approximate) AdS scale of the throat and by $L$
the size of the embedding manifold, whose precise geometry is again
not important. One has $L \gtrsim R_1$, since otherwise the throat
could not be glued into the manifold. If the embedding manifold is of
minimal size, $L \sim R_1$, KK modes with masses $\smash{m_n \ll
R_1^{-1}}$ cannot resolve its precise geometry.  We can then describe
the embedding manifold by the Planck brane in a Randall-Sundrum
model. Let us consider the Kaluza-Klein expansion of the graviton in
the throat. If we restrict ourselves to an s-wave with respect to the
compact manifold $\mathcal{M}$ multiplying the slice of AdS$_5$, we
can take the action from~\cite{Chung:2000rg} obtained in the context
of Randall-Sundrum phenomenology:\footnote{ The usual orbifold
boundary conditions were taken for the derivation of coupling
strengths and masses of graviton KK modes. It is not immediately clear
whether the same boundary conditions follow from a reduction to 5d of
a 10d geometry since the effective theory is defined on an interval
instead of an $\smash{S^1 / Z_2}$ orbifold. However, one can rederive
the Randall-Sundrum model on an interval if one takes Gibbons-Hawking
terms~\cite{Gibbons:1976ue} at the IR and the UV brane into
account. Varying with respect to the metric yields a condition similar
to the Israel junction condition, to be evaluated only at one side of
the brane. Inserting the background metric, one finds the relation
between the cosmological constants on the brane and in the bulk as
well as the usual boundary conditions for the fluctuations (see
e.g.~\cite{Chamblin:1999ya} for a derivation of the Israel junction
condition using Gibbons-Hawking terms).  }
\begin{equation}
\label{5daction}
S = \int d^4 x \; \sum_n \left( - \frac{1}{2} \partial_{\alpha} h_{\mu
\nu}^{(n)} \, \partial^{\alpha} h^{\mu \nu (n)} - \frac{1}{2} \, m_n^2
\, h_{\mu \nu}^{(n)} \, h^{\mu \nu (n)} + \frac{1}{\sqrt{2}}
\frac{g_n}{\sqrt{M_5^3 R_1}} \, h^{(n)}_{\mu \nu} \, T^{\mu \nu}
\right).
\end{equation}
The effective 5d Planck scale $M_5$ is determined by $M_5^3 \sim
M_{10}^8 R_1^5$.  We have included the coupling of the KK modes
$\smash{h_{\mu \nu}^{(n)}}$ to the energy-momentum tensor
$\smash{T^{\mu \nu}}$ on the Planck brane, which we will need in a
moment. For KK modes with $\smash{m_n \ll R_1^{-1}}$, the masses are
determined by
\begin{equation}
J_1(m_n/m_\IR) \simeq 0 \; \Rightarrow \; m_n \simeq \left( n
+\frac{1}{4} \right) \pi \, m_\IR,
\end{equation}
where $\smash{m_\IR = z_\IR^{-1}}$ is the inverse conformal length of
the throat (cf. Sect.~\ref{decayingaugepicture}) and we have used the
asymptotic form for large arguments of the Bessel function $J_1$. This
is consistent for $n$ somewhat larger than 1. The coupling constants
$g_n$ were calculated in~\cite{Chung:2000rg}, the result being
\begin{equation}
\label{gn}
g_n = \left( \left( \frac{Y_1(m_n R_1)}{Y_1(m_n/m_\IR)} \right)^2 -1
\right)^{-1/2} \simeq \; \sqrt{\frac{\pi}{2}} \, \sqrt{m_n m_\IR} \,
R_1.
\end{equation}
In the last step we have used the asymptotic forms for the Bessel
function $Y_1$.

Let us return to the case of two throats and consider another throat
in the embedding manifold. We take the throat to be AdS$_5
\times$S$^5$ such that it can be equally well described by a stack of
D3-branes. Again, its AdS scale $R_2$ cannot be larger than $L$, and
since we have assumed $L \sim R_1$, one has $R_1 \gtrsim R_2$. The
corresponding number $N_2$ of D3-branes follows from Eq.~\eqref{RN} as
$N_2 \sim M_{10}^4 R_2^4$. Now, when viewed from the first throat, the
gauge theory on the stack of $N_2$ D3-branes resides on the Planck
brane. Therefore, the graviton KK modes in this throat couple directly
to the energy-momentum tensor of the gauge theory. Using the last term
in Eq.~\eqref{5daction}, the decay of these KK modes into the other
throat can be calculated as a decay into gauge fields.\footnote{ There
are also decays into the fermions and scalars in the gauge
theory. However, the corresponding decay rates have the same order of
magnitude as the decay rate into gauge fields.  } By the standard
formula, the decay rate of a KK mode with mass $m_n$ into one species
of gauge fields is
\begin{equation}
\Gamma \sim \frac{g_n^2}{M_{10}^8 R_1^6} \, m_n^3.
\end{equation}
There are $N_2^2$ gauge fields in the adjoint representation of
U($N_2$).  Summing and using Eqs.~\eqref{gn} and~\eqref{RN}, the total
decay rate follows:
\begin{equation}
\label{DecayRate3}
\Gamma \sim \frac{R_2^8}{R_1^4} \, m_n^4 \, m_\IR.
\end{equation}
This result should be compared with Eq.~\eqref{DecayRate} from the
pure gauge theory calculation. The distance $A$ between the two
throats cannot be smaller than their AdS scales $R_1$ and $R_2$. Since
we have also assumed $L \sim R_1$ and $\smash{m_n \ll R_1^{-1}}$, the
second term in Eq.~\eqref{DecayRate} is dominant. Using $l = 0$ for
the s-wave that we have considered and $L \sim R_1$, we get the same
result as Eq.~\eqref{DecayRate3}, including the factor of $m_\IR$!

The above process is just the reverse of the energy loss by the heated
Planck brane considered, e.g., in~\cite{HMR,Langlois:2002ke}. Our
calculation can also be viewed as a rephrasing, using partly the gauge
theory picture and partly the gravity picture, of the tunneling
calculation performed in~\cite{smallnumbers}.  In these papers, the
decay rate of graviton KK modes between two throats was calculated in
a 5d model with two AdS$_5$ slices which are glued together at a
common Planck brane, assuming equal AdS scales $R_1=R_2$. However,
besides giving the corrections due to different AdS scales, from the
above derivation it is maybe more evident why the result is correct
also in a genuine 10d setup.

The decay rate $\smash{\Gamma \sim (m R)^4 \, m_\IR}$
from~\cite{smallnumbers} was used in a number of
papers~\cite{Barnaby:2004gg,Kofman:2005yz,Frey:2005jk,Chialva:2005zy}
in the context of reheating after brane-antibrane
inflation. Moreover,~\cite{Langfelder} contains a careful analysis in
a 5d model of effects related to the finite length of realistic
throats.  In this paper, the global KK modes in the two-throat system
are determined. Tunneling of KK modes is then viewed as the
decoherence of wave packets, which are set up in one throat. We have
used this picture in Sect.~\ref{decayingaugepicture}.

Tunneling in a compact 10d setup with throats was considered in~\cite{
Firouzjahi:2005qs,CT}. For the case $m^{-1} > L$, a decay rate of
$\smash{\Gamma \sim (m R)^{16} \, m_\IR}$ was derived, assuming that the
particle has to tunnel through two barriers described by the potential in
Eq.~\eqref{se}. We see a conceptual problem with this approach since we do not
know how to justify a 1-dimensional quantum-mechanical picture (this 1
dimension being the radial coordinate) in the two-throat case. But even if we
accept this description for the moment, there are further issues related to
the two-barriers assumption: The barriers extend to values of $r \sim
m^{-1}$ as can be seen from Fig.~\ref{pot}. Since $m^{-1} \gg R$ and
$r$ measures the physical distance for $r \gg R$
(cf.~Eq.~\eqref{threebrane}), the width of each barrier is given by
$m^{-1}$. This just reflects the fact that a particle with mass $m$
has a de Broglie wavelength of $m^{-1}$. Accordingly, the particle has
to tunnel through two entire barriers only if the distance $A$ between
the two throats is $\sim 2 m^{-1}$. Indeed, from Eq.~\eqref{DecayRate}
for $l=0$ and since $L > A$, we get a decay rate of $\smash{\Gamma
\sim (m R)^{16} \, m_\IR}$ in this case, in agreement
with~\cite{Firouzjahi:2005qs,CT}. However, if $A$ is smaller than
$\sim 2m^{-1}$, the particle has to tunnel through a smaller barrier.
Correspondingly, the decay rate becomes larger, as can be seen from
Eq.~\eqref{DecayRate}.

The case $m^{-1} \lesssim L$ (without assuming $m^{-1} \ll L$) was also
considered in~\cite{Firouzjahi:2005qs,CT}. It was found that the decay rate
can be much larger than $\smash{\Gamma \sim (m R)^{16} \, m_\IR}$ if a certain
resonance
condition is fulfilled. We have not determined the decay rate for this case
and therefore have no result to compare with.\footnote{Note that we have
assumed that $L \gg A,m^{-1}$ in deriving Eq.~\eqref{DecayRate2}. This result
is therefore not suitable to compare with the results from~\cite{
Firouzjahi:2005qs,CT} where the limit of extremely large $L$ was not taken.}
It would be interesting, though, to evaluate Eq.~\eqref{Matrixelement5} for
$m^{-1} \lesssim L$ and to see whether one can reproduce the results
from~\cite{Firouzjahi:2005qs,CT} as well as their resonance condition.

\section{Conclusions and Outlook}
\label{conclusions}
We have determined the energy loss rate $\dot{\rho}$ of a throat which
is heated to a certain temperature as well as the decay rate $\Gamma$
of single KK modes localized in a given throat. As a simplified setup
we have chosen a 6-dimensional torus with two AdS$_5 \times$S$^5$
throats. However, as we have argued in Sect.~\ref{energytransfer}, our
results stay parametrically correct for more general embedding
manifolds and throat geometries.  Especially, they are applicable for
two Klebanov-Strassler throats if the curvature of the space
connecting them is not larger than the inverse distance.

In earlier
investigations~\cite{smallnumbers,Barnaby:2004gg,Kofman:2005yz,Frey:2005jk,Chialva:2005zy,CT,Firouzjahi:2005qs,Langfelder}
of the decay of KK modes between throats, the decay/tunneling rate was
determined from solving wave equations in a given gravity
background. Most results were derived for the simple model of two
AdS$_5$ slices, glued together at a common Planck brane. As we have
explained in Sect.~\ref{comparison}, this calculation is difficult to
perform in a genuine 10d setup. Inspired by~\cite{K}, we instead chose
the dual gauge theory picture for our calculations. Namely, each
AdS$_5 \times$S$^5$ throat can be equally well described by a
corresponding stack of D3-branes. Both brane stacks are coupled by the
supergravity fields in the embedding space. The energy transfer rate
from a heated throat then follows from simple tree-level
quantum-field-theory processes. For the decay rate of throat-localized
KK modes which are dual to glueballs, we first had to determine the
glueball-supergravity vertex. To this end, we have calculated the
decay rate of throat-localized KK modes into flat 10d space in the
gravity picture. Then, we have determined the glueball-supergravity
vertex by demanding that the decay rate following from this vertex
give the same result.  We have also presented some cross-checks
from the gravity picture in Sect.~\ref{comparison}.

From our analysis, we were able to determine the dependence of the
energy transfer and decay rates on the distance $A$ between two
throats as well as on the size $L$ of the embedding manifold.  For
example, this is relevant for the analysis of reheating after
brane-antibrane inflation.  In such models, one often considers
inflation occurring in one throat, whereas the standard model branes
reside at the bottom of another, longer throat. In that way, the
generation of the right level of density fluctuations is reconciled
with a solution of the gauge hierarchy problem \`a la
Randall-Sundrum. For a viable reheating of the standard model sector,
it is crucial that the energy from brane-antibrane annihilation is
transferred efficiently into the standard model throat. This
question was analysed
in~\cite{Barnaby:2004gg,Kofman:2005yz,Frey:2005jk,Chialva:2005zy,CT}.
We find that, as long as the embedding manifold is not of minimal
size, the energy transfer rate Eq.~\eqref{el} is considerably lower
than the rates previously derived
in~\cite{Barnaby:2004gg,Kofman:2005yz,Frey:2005jk,Chialva:2005zy}. Given our
results, it will be interesting to reconsider reheating after brane-antibrane
inflation.

Our results remain applicable if one deals with a small stack of
D3-branes.\footnote{ The sole exception is the decay rate of
brane-localized states. In this case, our derivation of the vertex
from the gravity picture does not work, since supergravity is not a
good approximation.}  An interesting setup is the following: Consider
that the standard model resides on some D-branes in a given Calabi-Yau
orientifold. Since they are a common feature of flux
compactifications, such a manifold will typically contain several
throats~\cite{Hebecker:2006bn}.  Modelling the standard model branes
by a small stack of D3-branes, we can estimate the rate of energy loss
to the throats in early cosmology from Eq.~\eqref{el}. According to
Eq.~\eqref{RN}, with $N$ being small, we just have to replace one
factor of $\smash{R^8}$ by the corresponding power of the 10d Planck
scale, $\smash{M_{10}^{-8}}$. The throat sectors, which are heated up
in that way, may provide interesting dark matter
candidates~\cite{Kofman:2005yz,CT}. Later in cosmological evolution,
throat-localized KK modes may decay back to the standard model. The
corresponding rate can be estimated from Eq.~\eqref{DecayRate}, again
replacing $\smash{R_2^8}$ by $\smash{M_{10}^{-8}}$. The decay rate
strongly depends on the angular momentum of the throat-localized KK
modes. We have given this dependence explicitly for the angular
momentum with respect to an S$^5$ in an AdS$_5 \times$S$^5$
throat. Moreover, we have outlined how to determine this dependence
for other manifolds, e.g.~the (approximate) T$\smash{^{1,1}}$ in a
Klebanov-Strassler throat.  Depending on the cosmological epoch, the
decaying KK modes may influence the abundances of light elements or
lead to diffuse gamma-ray background radiation, both effects being
strongly constrained by observations (see e.g.~\cite{UP}). Along these
lines it may even be possible to impose certain phenomenological
constraints on multi-throat compactifications.

More generally, one may discuss several cosmological scenarios where
reheating takes place either in the standard model sector or in a
throat (as is the case after brane-antibrane inflation) and the
standard model resides either at the bottom of a throat or somewhere
in (the rest of) the Calabi-Yau orientifold. The energy transfer and
decay rates that we have calculated can then be used in a set of
Boltzmann equations to determine the evolution of energy densities of
the standard model and throat sectors. We leave these interesting
applications for future work.

\noindent
{\bf Acknowledgements}:\hspace*{1cm}We would like to thank J.~Braun,
F.~Br\"ummer, X.~Chen, D.~Dietrich, L.~Kofman and M.~Trapletti for helpful
comments and discussions.

\appendix
\section*{Appendix: Evaluation of the propagator}
In Eq.~\eqref{propagator}, we had to evaluate the following propagator
in a mixed, energy-configuration-space representation:
\begin{equation}
\label{appendixgl}
\int \frac{d^6 \rho}{(2 \pi)^6 } \, \frac{ e^{ i \vec{A} \,
\vec{\rho}}}{m^2 -\vec{\rho}^2 +i \epsilon}.
\end{equation}
We perform the integral for imaginary values $m \rightarrow e^{i
\pi/2} m$ and use analytic continuation. The integral changes into
\begin{equation}
- \int \frac{d^6 \rho}{(2 \pi)^6 } \, \frac{ e^{ i \vec{A} \,
  \vec{\rho}}}{m^2 +\vec{\rho}^2 }.
\end{equation}
We can then employ the identity $\smash{c^{-1}=\int_0^\infty d \tau
e^{-c \tau}}$ for $\Re{c} > 0$ and get
\begin{equation}
\begin{split}
& \frac{-1}{(2\pi)^6} \int_0^\infty \hspace{-1.5mm} d \tau
\hspace{-.5mm} \int \hspace{-1mm} d^6 \rho \;\; e^{ i \vec{A} \,
\vec{\rho}} \, e^{ -(m^2 +\vec{\rho}^2) \tau } \\ = & \frac{-1}{(2
\pi)^6} \int_0^\infty \hspace{-1.5mm} d \tau \, \left( \left[ \int
\hspace{-1mm} d \rho_1 \; e^{ i A_1 \, \rho_1} \, e^{ -\rho_1^2 \tau }
\right] \cdots \left[ \int \hspace{-1mm} d \rho_6 \; e^{ i A_6 \,
\rho_6} \, e^{ -\rho_6^2 \tau } \right] e^{ -m^2 \tau} \right) \\ = &
\frac{-1}{(4 \pi)^3} \int_0^\infty \hspace{-1.5mm} d \tau \; \frac{1}{
\tau^3} \, e^{-A^2 / 4 \tau } \, e^{-m^2 \tau}.
\end{split}
\end{equation} 
We have used that $A^2 = A_1^2 + \cdots + A_6^2$. According to
Eq.~3.471.9 in \cite{GR}, this integral can be evaluated in terms of
the modified Bessel function $K_{-2} \equiv K_2$, which yields 
\be
\frac{-1}{(2\pi)^3} \frac{m^2}{A^2} \, K_2( m A).  
\ee 
Following from
Eq.~9.6.4 in \cite{AS}, $K_2$ is related to the Hankel function
$H_2^+=J_2+i Y_2$. The above expression can be written as 
\be
\frac{i}{(4\pi)^2} \frac{m^2}{A^2} \, H^+_2( e^{i \pi/2} m A).
\ee
The Hankel function has a branch cut along the negative real
axis. Therefore, one can analytically continue back to real values $m
\rightarrow e^{-i \pi/2} m$, which gives
\begin{equation}
\label{resultappendix}
\frac{-i}{(4\pi)^2} \frac{m^2}{A^2} \, H^+_2( m A).
\end{equation}

\end{document}